\documentclass[%
 prb,
 aps,
 amsmath,amssymb,
 reprint,%
 showpacs,
 floatfix,
]{revtex4-1}

\usepackage{graphicx}
\usepackage{dcolumn}
\usepackage{bm}
\usepackage{mathtools}

\begin{document}


\title{Majorana modes in smooth normal-superconductor nanowire junctions} 



\author{Javier Osca}
\email{javier@ifisc.uib-csic.es}
\affiliation{Institut de F\'{\i}sica Interdisciplin\`aria i de Sistemes Complexos
IFISC (CSIC-UIB), E-07122 Palma de Mallorca, Spain}
\author{Lloren\c{c} Serra}
\affiliation{Institut de F\'{\i}sica Interdisciplin\`aria i de Sistemes Complexos
IFISC (CSIC-UIB), E-07122 Palma de Mallorca, Spain}
\affiliation{Departament de F\'{\i}sica,
Universitat de les Illes Balears, E-07122 Palma de Mallorca, Spain}

\date{July 26, 2013}

\begin{abstract}
A numerical method to obtain the spectrum of smooth normal-superconductor junctions
in nanowires,
able to host Majorana zero modes, is presented.
Softness in the potential and superconductor interfaces yields opposite effects on the 
protection of Majorana modes. While a soft potential is a hindrance
for protection, a soft superconductor gap transition greatly favors  it.  
Our method also points out the possibility of extended Majorana states
when propagating modes are active far from the junction, although
this requires equal incident fluxes in all open channels.
\end{abstract}


\maketitle 
\section{Introduction}

In 1936 Ettore Majorana theorized the existence of elementary particles,
now called Majorana Fermions, that coincide 
with their own antiparticles.\cite{Majorana}
The implementation of quasiparticle excitations having a similar property,
called Majorana states, is currently attracting much interest in condensed matter systems in general,\cite{Kitaev,Wilczeck,Alicea,Qi,Leijnse,Beenakker,StanescuREV,Fu,Akhmerov,Tanaka,Law} 
and in nanowires in particular.\cite{Lutchyn,Oreg,Stanescu,Flensberg, Potter,Potter2,Ganga,Egger,Zazunov,Klino1,Klino2,Lim,Lim2,Lim3} 
Interest has been further fueled by recent experimental evidences of these Majorana states in quantum wires.\cite{Mourik,Deng,Rokhinson,Das,Finck} 
They are also known as Majorana zero modes (MZM) and, in essence, 
they are topological zero-energy states living close to the 
system edges or interfaces. 
The existence of an energy gap between the MZM and nearby excitations protects the former from decoherence. These properties make MZM's interesting 
not only for their exotic fundamental physics but also for their
potential use in future topological quantum-computing 
applications.\cite{Pachos,Nayak}

Majorana modes can be implemented in a superconductor wire
by the combined action
of superconductivity, Rashba spin-orbit coupling and Zeeman magnetic effect. 
In a superconductor, nanowire electrons play the role of particles, while holes
of opposite charge and spin perform the role of antiparticles. Superconductivity leads to a charge symmetry breaking and allows quasiparticles without a good isospin number. On the other hand, the Rashba effect is a direct result of an inversion asymmetry caused by an electric field in a direction perpendicular to the propagation while the Zeeman magnetic field breaks the spin rotation symmetry of the system. The combined action of both couplings can create effective spinless 
Majorana states.

It is known that in semiconductor nanowires having a region of
induced superconductivity
Majorana edge states are formed in the junction between the superconductor and the normal side of the nanowire. 
This work addresses the
physics of soft-edge junctions, where both superconductivity and potential 
barrier characterizing the edge vary smoothly as one moves from the normal to the superconductor side.
Previous works on nanowire Majorana physics assumed abrupt
transitions, with the exception of Ref.\ \onlinecite{Kells} that considered a linear
potential edge in a constant superconductivity. 
Our work generalizes
Ref.\ \onlinecite{Kells} by allowing also a diffuse superconductivity edge placed
either at the same or at different position of the potential barrier.
We find a strong influence of the superconductivity smoothness on the 
finite-energy Andreev states occurring inbetween potential and superconductivity 
edges. This is relevant for the protection of MZM's as it is affected 
in opposite ways by the smoothness in potential and superconductivity:
protection is hindered by a smooth potential (also discussed 
in Ref.\ \onlinecite{Kells}) and, remarkably, it is favored by a smooth 
superconductivity.

This work is divided
in six sections. In Sec.\ II the physical system is introduced and 
in Sec.\ III the numerical method is explained. 
Section IV contains different results for bound and resonant states in
different kinds of junctions, in absence of any input flux. 
Section V is devoted to an extension for junctions under an input 
flux, demonstrating that in this case extended MZM's
are possible, as opposed to 
the localized ones of preceding sections. 
The conclusions are drawn in 
Sec.\ VI.

\section{Physical system}

We consider a purely 1D nanowire model with spin orbit interaction inside an homogenous 
Zeeman magnetic field as in Ref.\ \onlinecite{Serra}. 
The system is described by a Hamiltonian
of the 
Bogoliubov-deGennes kind,  
 \begin{eqnarray}
	\mathcal{H}_{\it BdG} &=& 
	\left( \frac{p_{x}^{2}}{2m} + V(x) -\mu \right)\tau_z 
	+ \Delta_B\, \vec{\sigma}\cdot\hat{n}\nonumber\\
	&+& \Delta (x)\, \tau_{x}+\frac{\alpha}{h}\, p_x \sigma_{y}\tau_z\;,
	\label{E1}
\end{eqnarray}
where
the Pauli operator for spin is represented by $\vec{\sigma}$ while the operator for isospin (electron/hole charge) is represented
by $\vec{\tau}$.  The successive energy contributions in Eq.\ (\ref{E1}) are (in left to right order): kinetic, electric potential, chemical potential, Zeeman, superconduction and Rashba term. The latter arises from the self interaction between an electron (or hole) spin with its own motion due to the presence of a transverse electric field, perceived as an effective magnetic field in the rest frame of the quasiparticle. On the other hand, the Zeeman effect is the band splitting caused by the application of an external magnetic field. Rashba spin orbit and Zeeman effects depend on the parameters $\alpha$ and $\Delta_B$ respectively. 
Since we consider a nanowire made of an homogenous material inside a constant magnetic field these parameters are assumed homogenous. The magnetic field 
points in the $\hat{x}$ direction, parallel to the propagation direction and perpendicular to the spin orbit effective magnetic field direction $\hat{y}$. 
The superconduction term arises from a mean field approximation over the phonon assisted attractive interaction between electrons. This leads to the coupling of the opposite states of charge of the base and the creation of Cooper pairs whose breaking energy is the energy gap $\Delta(x)$. The remaining terms in Eq.\ (\ref{E1}) are the potential term $V(x)$ created by the presence of a metallic gate over the nanowire and the chemical potential term $\mu$.

\begin{figure}[t]
\centering
\includegraphics[width=7.5cm]{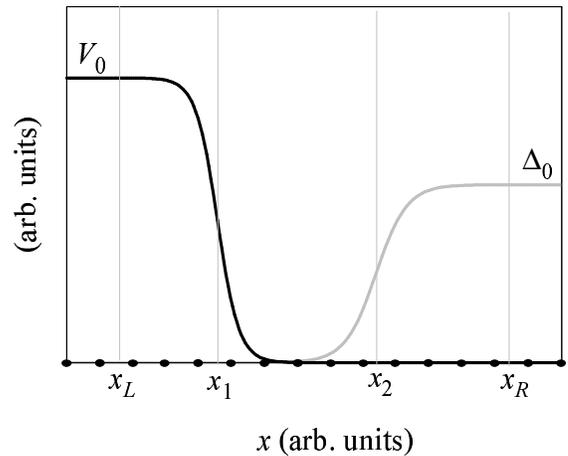}
\caption{NS junction of an infinite nanowire. The black curve is the potential $V(x)$ created by a gate in contact with
the nanowire while the gray curve is the superconductor gap induced by proximity with an s-wave superconductor. The normal contact ($x<x_L$) is characterized by potential $V_0$ 
and the superconductor one ($x>x_R$) by a gap $\Delta_0$. A smooth variation of 
$V(x)$ and $\Delta(x)$ occurs at transition points $x_1$ and $x_2$, respectively.
A Zeeman magnetic field is applied homogenously along the entire nanowire pointing in 
$x$-direction while the Rashba SOI effective magnetic field points perpendicularly in $y$-direction.
The numerical method uses a grid as indicated schematically by the dots on the $x$ axis.}
\label{F1}
\end{figure}

The nanowire smooth junction is sketched in Fig.\ \ref{F1}, with
left ($x<x_L$) and right ($x>x_R$) contacts corresponding 
to the normal and superconductive sides, respectively. 
The normal contact is characterized by a bulk potential $V_{0}$ and 
the superconducting one by a 
gap $\Delta_0$. 
Superconduction in a semiconductor nanowire region is achieved by maintaining that region in contact with a 3D superconductor. 
In the junction region between the two asymptotic behaviors, $x_L < x < x_R$,
a smooth transition is described by the potential
$V(x)$ and  gap $\Delta(x)$ functions of the position $x$. 

The transitions between bulk values in $V(x)$ and $\Delta(x)$ 
are modeled with two soft Fermi functions centered at $x_1$ and $x_2$, respectively. Their 
softness is controlled with parameters $s_1$ and $s_2$. 
A zero softness means a step interface, while a high value implies a smooth one. 
These two functions read
\begin{eqnarray}
	\label{E1b}
	V(x) &= & \frac{V_0}{1+e^{(x-x_1)/s_1}}\;, \\
	\label{E1c}
	\Delta(x)&=& \Delta_0\left(1-\frac{1}{1+e^{(x-x_2)/s_2}}\right)\;.
\end{eqnarray}

\section{Numerical method}

The energy eigenstates fulfill the time independent Schr\"odinger equation
with the Bolgoliubov-deGennes Hamiltonian,
 \begin{equation}
	\left( \mathcal{H}_{BdG}-E\right) \,\Psi(x,\eta_{\sigma},\eta_{\tau})
	=0\;,
	\label{E2}
\end{equation}
where the wave function variables are the spatial coordinate 
$x\in (-\infty,\infty)$, the spin $\eta_{\sigma}\in \{\uparrow,\downarrow \}$ and
the isospin $\eta_{\tau}\in \{ \Uparrow,\Downarrow \}$. The basis projection 
for spin and isospin 
is taken in $\hat{z}$ orientation, with isospin up and down representing
electron and hole quasiparticles, respectively. We expand next the wave function
in spin and isospin spinors
\begin{equation}
	\Psi(x,\eta_{\sigma},\eta_{\tau})=\sum_{s_\sigma s_\tau}{\Psi_{s_\sigma s_\tau}(x)\,\chi_{s_\sigma}(\eta_{\sigma})\,\chi_{s_\tau}(\eta_{\tau})}\;,
	\label{E3}
\end{equation}
with the quantum numbers $s_\sigma=\pm$ and $s_\tau=\pm $. 
The spin and isospin states fulfill
\begin{eqnarray}
	\vec{\sigma}\cdot\hat{n}\, \chi_{s_\sigma}(\eta_{\sigma}) &=& s_{\sigma} \chi_{s_\sigma}(\eta_{\sigma})\,,
	\label{E5}\\
	\tau_{z}\, \chi_{s_\sigma}(\eta_{\sigma}) &=& s_{\tau} \chi_{s_\tau}(\eta_{\tau})\,.
	\label{E6}
\end{eqnarray}

We numerically obtain the wave function amplitudes  
$\Psi_{s_\sigma s_\tau}(x)$ on the set of $N$ grid points, qualitatively sketched in Fig.\ \ref{F1}
($N$ is actually much larger than shown in the figure). In our approach, the energy $E$ is given 
and we determine whether a physical solution exists or not for that energy. 
In particular, MZM's will be found for values of $E$ equal to zero.
Using $n$-point finite difference formulas for the $x$-derivatives, Eq.\ (\ref{E2}) transforms on the grid 
into a matrix linear equation of homogenous type. 

The solution must be compatible with the bulk boundary conditions for grid points in
the normal $(x<x_L)$ and superconductor contacts $(x>x_R)$. In these asymptotic regions 
the solutions, at the desired energy $E$, are given by 
a linear combination of bulk eigensolutions $\Phi_{k}^{(c)}(x,\eta_{\sigma},\eta_{\tau})$, each one characterized by a wave number $k$ and $c=L,R$ being a generic label for the contact,
 \begin{equation}
	\Psi(x,\eta_{\sigma},\eta_{\tau})=\sum_{k}{C_k^{(c)}\, \Phi_{k}^{(c)}(x,\eta_{\sigma},\eta_{\tau})}\,.
	\label{E7_2}
\end{equation}
The bulk eigensolutions  are expressed in terms of  exponentials
 \begin{equation}
	\Phi_{k}^{(c)}(x,\eta_{\sigma},\eta_{\tau})=\sum_{s_\sigma s_\tau}{\Phi_{k s_\sigma s_\tau}^{(c)} e^{ik(x-x_c)}\chi_{s_\sigma}(\eta_{\sigma})\chi_{s_\tau}(\eta_{\tau})}\,.
	\label{E8}
\end{equation}

The set of wave numbers and state coefficients $\{k,\Phi_{k s_\sigma s_\tau}^{(c)}\}$ characterizing 
the solutions in contact $c$ must be known in advance in order to proceed with the numerical calculations. These coefficients can be obtained 
for the homogenous and infinite problem either
analytically or by means of additional numerical methods.\cite{Oreg,Serra} 
Equation (\ref{E7_2}) must be fulfilled
in replacement of Eq.\ (\ref{E2}) for grid points in the asymptotic regions. Notice 
that they are local relations in $x$ and, therefore, do not involve wave function amplitudes on 
points located further to the left or right of the grid ends.

Due to symmetries, there are always four bulk wave numbers per contact in {\em outward} direction.
By {outward} direction we mean either exponentially decaying from the junction, in case of
evanescent modes, or moving away from it in case of propagating modes. Notice that for propagating modes the flux direction is parallel and antiparallel to the corresponding real $k$ for quasiparticles
of electron and hole type, respectively.
A closed linear system for the set of $4N+4+4$ unknowns 
$\{\Psi_{s_\sigma s_\tau}(x), C^{(L)}_k, C^{(R)}_k  \}$ is easily obtained from
Eqs.\ (\ref{E2}) and (\ref{E7_2}). A final complication, however, is found in the homogenous
character of this linear system mathematically admitting the trivial solution of all unknowns 
equal to zero.

We discard the trivial solution by introducing an arbitrary matching point $x_m$ 
as well as a specific pair of spin-isospin
components $(s,t)$. Assuming $\Psi_{st}(x_m)$ does not identically vanish we can arbitrarily 
impose
\begin{eqnarray} 
\Psi_{st}(x_m) &=& 1\; , \label{E9}\\
\!\!\!\!\!\!\!\!\!\!\!
\left(\frac{d^{(L)}}{dx}-\frac{d^{(R)}}{dx} \right) \Psi_{s_\sigma s_\tau}(x_m)&=&0\;,  \quad (s_{\sigma},s_{\tau})\neq (s,t)\,.
			\label{E10}
\end{eqnarray}
Equations (\ref{E9}) and (\ref{E10}) are four equations that we require at $x_m$ in place of the 
Bogoliubov-deGennes one. Thanks to Eq.\ (\ref{E9}) the resulting system is no longer 
homogenous. 
In Eq.\ (\ref{E10}), $d^{(L)}/dx$ and $d^{(R)}/dx$ indicate grid derivatives using only left or right 
grid neighbors. Crossing the matching point is actually avoided using non centered
finite difference formulas. With this substitution of one equation the resulting linear system
admits a nontrivial solution, robust with respect to changes in the arbitrary choices: $x_m$, $(s,t)$. 

By means of  
Eq.\ (\ref{E10}) our algorithm 
ensures the continuity at the matching point of the first derivative for all spin-isospin
components, with the exception of the arbitrarily chosen $(s,t)$. This relaxation of one condition
makes the algorithm numerically robust and free from singularities. The mathematical solutions
can be discriminated by defining the physical {\em measure} 
\begin{equation}
	\mathcal{F}=\left| \left(\frac{d^{(L)}}{dx}-\frac{d^{(R)}}{dx}\right)
	\Psi_{st}(x_m) \right|^2\,.
	\label{E11}
\end{equation}
Only those results with ${\cal F}=0$ are true physical solutions but this can be tested afterwards,
at the end of the algorithm. Varying the energy $E$ or the Hamiltonian parameters
the method allows the exploration of the topological phases.

The resulting system of equation is solved with a sparse-matrix linear algebra package.\cite{Harwell}
As in Refs.\ \onlinecite{Oreg} and \onlinecite{Serra} the numerical algorithm works in adimensional units, using the Rashba spin-orbit interaction (SOI) as a reference. The corresponding length and energy units  read
\begin{eqnarray}
	L_{\it so}&=&\frac{\hbar^2}{\alpha m}\,,\\
	E_{\it so}&=&\frac{\alpha^2 m}{\hbar^2}\,.
	\label{E11_2}
\end{eqnarray}

\section{Results without input flux}

We study first  the physics of the junction in absence of any input 
fluxes. Physically, this situation occurs when propagating modes in both contacts are either not active
or, at most,  they carry flux only in outwards
direction from the junction. This behavior is expected in presence of purely absorbing
(reflectionless) contacts.
It is well known that in absence of propagating modes bounded MZM's may 
exist in some cases. 
The allowed asymptotic wave numbers have an imaginary component 
causing the wave functions to decay away from the junction. 
As a consequence the main characteristic of these bounded MZM's is that they are confined in a particular region of space. We will first check our method comparing with  the analytical 
limits of Klinovaja and Loss,\cite{Klino2} extending later the analysis to other results not obtainable analytically. These results range from the  formation of Majorana modes in soft edge junctions of different kinds, to the influence of the edge on the MZM localization and 
protection.

\subsection{Comparison with analytical expressions}	
\begin{figure}[t]
\centering
\includegraphics[width=8.5cm]{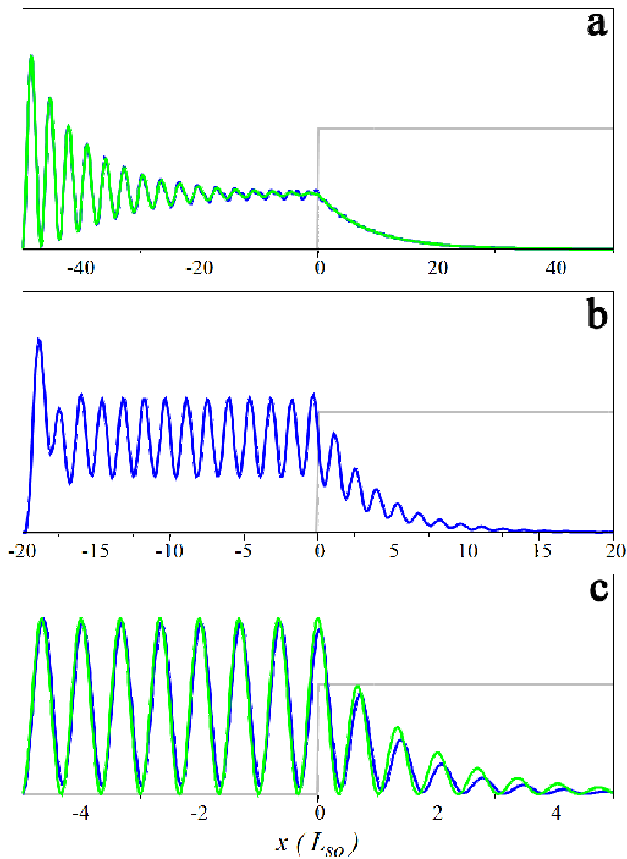}
\caption{
(Color online)
Density distributions of MZM's obtained with our 
numerical method (dark gray or blue) and
with analytical approximations (light gray or green).\cite{Klino2} 
a) Strong SOI regime $E_{so}>>\Delta_B$: $\Delta_B=0.09E_{so}$, $\Delta_0=0.06E_{so}$ and $L=50L_{so}$.  
b) Intermediate regime $E_{so}=\Delta_B$: $\Delta_0=0.2E_{so}$, $L=20L_{so}$.
In this case the analytical result is not known.
c) Weak SOI regime $E_{so}<<\Delta_B$: $\Delta_B=10E_{so}$, $\Delta_0=4E_{so}$ and $L=5L_{so}$.}
\label{F3_1}
\end{figure}

Reference \onlinecite{Klino2} provides analytical expressions for MZM's in a sharp NS junction, in a semi-infinite system. They are approximations valid deep into the topological phase $\Delta_B\gg\sqrt{\Delta_o^2+\mu^2}$. 
The approximations are done for both strong SOI ($E_{so}>>\Delta_B$) 
and weak SOI ($E_{so}<<\Delta_B$) regimes. In the strong SOI regime the Rashba spin orbit effect is the dominating term while the magnetic field and the superconductivity are treated as small perturbations. On the other hand, in the weak SOI regime the magnetic field term dominates. In Fig.\ \ref{F3_1} density distributions for NS junctions in a semi-infinite nanowire are shown for the strong and weak SOI regimes, as well as for an intermediate situation. The strong and weak 
regime numerical solutions (in dark-blue) are compared with their
analytical counterparts (in light-green). The exclusion effect 
on the hard edge on the left
is achieved 
in the numerical method 
by putting a very high sharp potential step at $x=-L$, while the 
sharp superconductor interface is located at $x=0$.

The strong SOI Majorana density function is characterized by the combination of an oscillatory behavior modulated by exponential bounds in the normal side of the junction while, on the other hand, the weak SOI density is characterized by constant oscillations up to the NS interface. Entering the superconductor contact
both densities decay, although in a more oscillatory way for the weak SOI. 
Note also that the intermediate regime $E_{so}\approx\Delta_B$ represents a sort of mixed situation with a first density peak near the $x=-L$ edge 
followed by regular oscillations of constant amplitude up to the NS junction. The theoretical and numerical results agree well in their corresponding regimes (but for some fine effects). However, the analytical solutions are not applicable out of their regimes of approximation. Therefore a numerical approach is potentially very useful in order to predict MZM's density distributions in many realistic physical realizations that can be out of the strong and weak regimes in a varying degree.

\subsection{Soft edge junction results}

\begin{figure}[t]
\centering
\includegraphics[width=8.5cm]{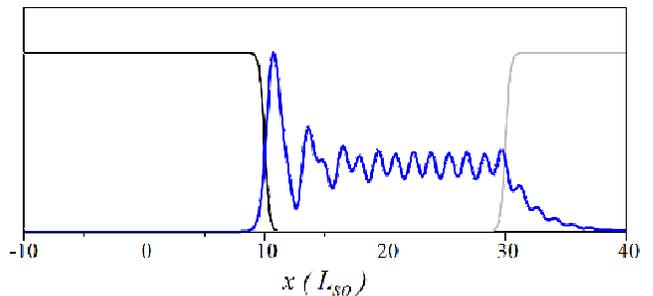}
\caption{(Color online)
MZM density (thick curve) in arbitrary scale when
the potential interface is located at $x_1=10L_{so}$ with a softness parameter $s_1=0.2L_{so}$ and the superconductor gap interface is located at $x_2=30L_{so}$ with softness $s_2=0.2L_{so}$. The rest of parameters are $V_o=2E_{so}$, $\Delta_o=0.25E_{so}$, $\Delta_B=0.4E_{so}$ and $\mu=0.1E_{so}$.
The position dependent potential and superconductor gap are shown by the thin
black and gray curves, respectively. }
\label{F3_2}
\end{figure}

Assume now the normal side contains a soft potential step
characterized by a 
finite $V_0$, allowing some penetration. 
As can be seen in Fig.\ \ref{F3_2},
this implies the appearance of a maximum in the density distribution near the potential edge followed by regular oscillations of decreasing amplitude. 
The density starts decaying exponentially in the superconductor interface until it vanishes
well inside the superconductor side of the system.

\begin{figure}[t]
\centering
\includegraphics[width=8.5cm]{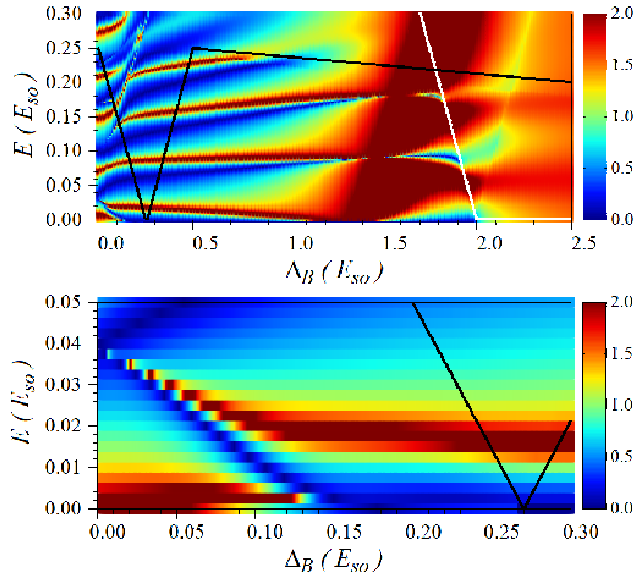}
\caption{
(Color online)
a) Junction spectrum  
for different values of the magnetic field. 
The same parameters of Fig.\ \ref{F3_2} have been used. 
The colors represent the values of the function $\mathcal{F}$. Note that a solution exists when $\mathcal{F}=0$ (in dark-blue). 
Above the black line propagating modes exist in the superconductor side of the junction while above the white line the propagating modes exist in the normal side. 
b) Zoom of the spectrum showing 
the formation of the Majorana zero mode when the magnetic field 
becomes high enough  for the system to perform a transition to the topological phase. $\Delta_B=0.27E_{so}$ is the topological critical value of the Zeeman field.}
\label{F3_3}
\end{figure}

The present method allows us to obtain the solutions not only for $E=0$ but for any arbitrary value of $E$. Figure \ref{F3_3} shows the location of the 
eigenstates in a $\Delta_B$-$E$ plane.
They are signaled by the zeros of the function $\mathcal{F}$ [cf.\ Eq.\ (\ref{E11})], represented here in a color (gray scale) plot.  
Black and white curves in Fig.\ \ref{F3_3} inform us on the presence of propagating modes in the superconductor and normal sides, respectively. 
That is, for energies above the curve propagating modes are possible in the
superconductor (black) and normal (white) contacts.
When propagating modes become possible asymptotically,
the zeros of $\mathcal{F}$ no longer represent
bounded states, but
purely outgoing resonances created by the 
junction.
In this case there is no violation of charge conservation since outgoing 
electron and hole equal fluxes imply zero currents.

For Zeeman energies lower than the critical value 
$\Delta_B^{(c)}\equiv\sqrt{\Delta_0^2+\mu^2}$
no MZM exists but, instead, finite energy subgap fermions may be found. 
Only those at positive energies are shown in Fig.\ \ref{F3_3}, 
but the spectrum is exactly symmetrical for negative energies. 
When the magnetic field energy equals $\Delta_B^{(c)}$
the gap closes in the superconductor side.
This is hinted in Fig.\ \ref{F3_3} by the presence of propagating modes in the superconductor side of the junction even at zero energies for this specific magnetic
field.
For higher fields the gap immediately reopens in the supercoductor region
and the junction enters
the topological phase with an $E=0$ solution, a MZM. In this phase
finite energy resonant Andreev states can be found as well. The energy difference 
between the MZM and the finite energy
states is a measure of the protection of the MZM. The greater the energy difference the greater the protection of the Majorana. 
Increasing further the magnetic field
the MZM is finally destroyed due to the closing of the gap in 
the normal side of the junction. 
This is signaled by the appearance of propagating modes in this side of the junction even at zero energy. When the state at zero energy becomes propagating the bounded Majorana zero modes can not exist. All these results are in agreement with the present knowledge on  MZM's and represent a further check on our numerical method.

\subsection{Softness effects}
\begin{figure}[t]
\centering
\includegraphics[width=8.5cm]{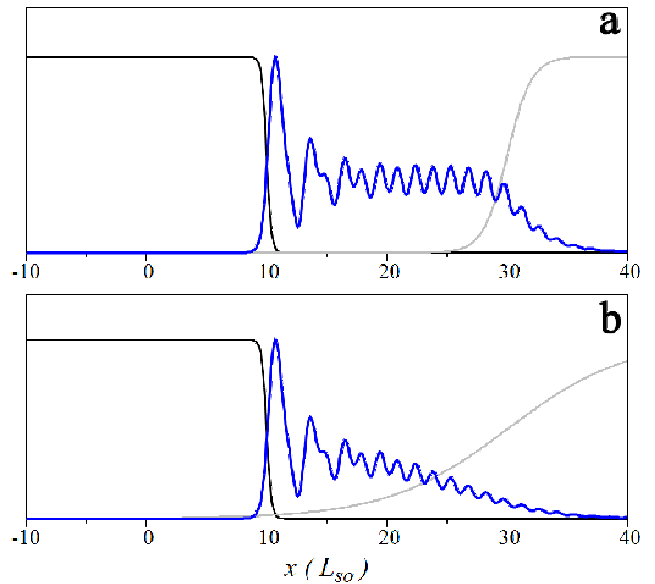}
\caption{
(Color online)
a) Same as in Fig. \ref{F3_2} but with a softer superconductivity interface $s_2=L_{so}$. b) Same as in Fig. \ref{F3_1} and a) but with an even softer, almost linear, superconductivity interface $s_2=5L_{so}$.}
\label{F4}
\end{figure}

This section is devoted to study the effects that changes 
in the softness parameter of the potential and superconductor interfaces
cause on the Majorana density function and on the junction spectrum. 
In general, the shape of the density function is robust to moderate changes in the softness of the superconductor gap interface (see Fig.\ \ref{F4}a). If we replace the sharp superconductor interface by a region of a gradually increasing superconductivity the Majorana wave function is not greatly affected. However, if we assume a very soft (almost linear) increase in superconductivity it is possible to see how the 
density tail of the MZM adapts to the appearance of the 
superconductivity (see Fig.\ \ref{F4}b).

\begin{figure}[t]
\centering
\includegraphics[width=8.5cm]{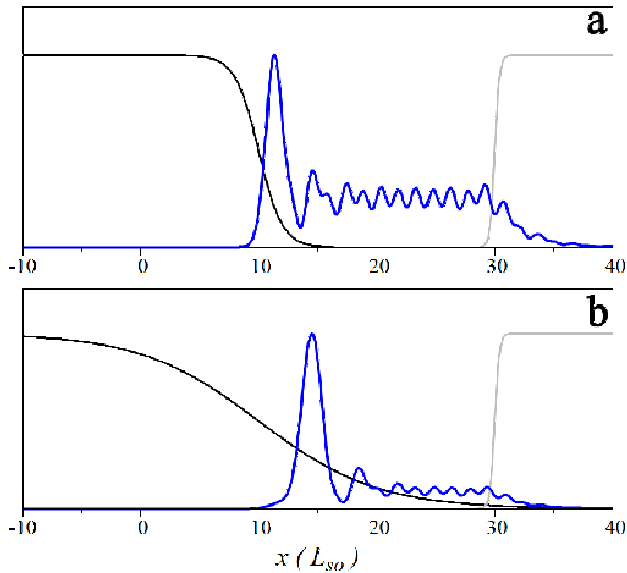}
\caption{
(Color online)
a) Same as in Fig.\ \ref{F3_2}  but with a potential softness parameter $s_1=L_{so}$ while the superconductor gap interface has a softness parameter $s_2=0.2L_{so}$. b) Same as a) but with a potential softness parameter $s_1=5L_{so}$.}
\label{F5}
\end{figure}

\begin{figure}[t]
\centering
\includegraphics[width=8.5cm]{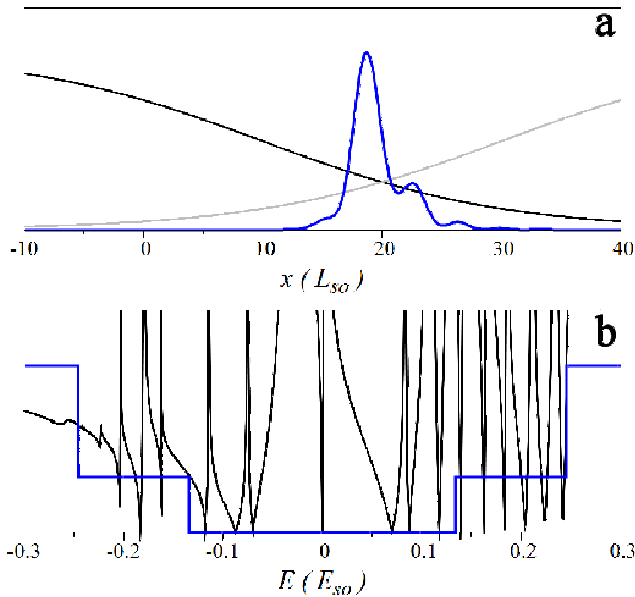}
\caption{
(Color online)
a) MZM density 
when the potential interface is located at $x_1=10L_{so}$ while the superconductor interface is located at $x_2=30L_{so}$, both with a softness parameter $s_1=s_2=10L_{so}$. 
The rest of parameters are $V_o=2E_{so}$, $\Delta_o=0.25E_{so}$, $\Delta_B=0.4E_{so}$ and $\mu=0.1E_{so}$. b) Energy spectrum of the junction in a). As before the function $\mathcal{F}$ is shown in black while the number of propagating modes in the superconductor side of the junction is shown in blue-gray. Zeros in $\mathcal{F}$ indicate the existence of a solution with the corresponding energy $E$.} 
\label{F6}
\end{figure}

The MZM is also robust with changes of the potential interface softness 
(see Fig.\ \ref{F5}a). Again, only with an almost linear decrease of the junction  
potential a sizeable reduction of the density tail towards the supercoductor side 
can be seen with respect 
to the result for an abrupt potential.
We also notice a slight increase in the width of the density peak as
well as a change in the peak position (see Fig.\ \ref{F5}b).
Combining the two effects, if the softness of both potential and superconducting interfaces is high enough a MZM with a well 
localized density peak is found (see Fig.\ \ref{F6}a).

\begin{figure}[t]
\centering
\includegraphics[width=8.5cm]{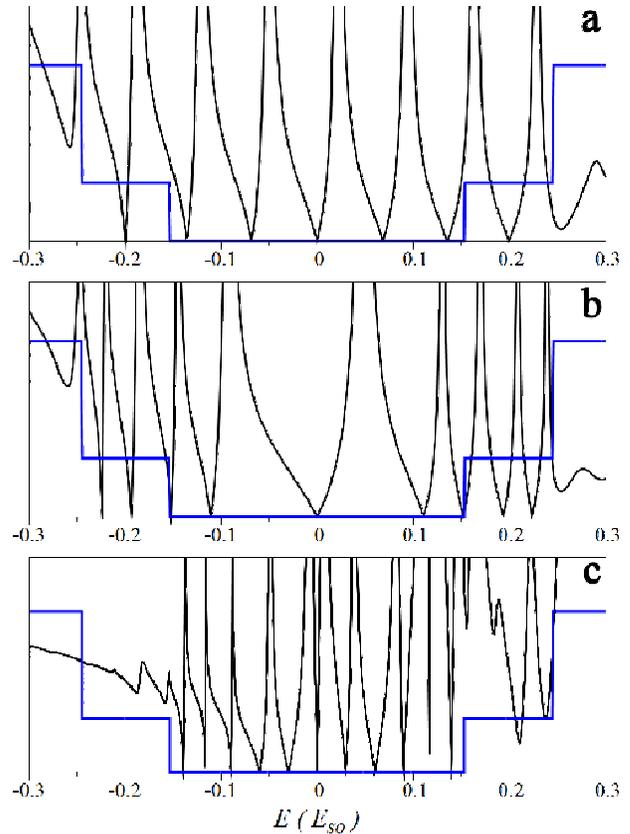}
\caption{
(Color online)
a) Junction spectrum when the potential interface is located at $x_1=10L_{so}$ while the superconductor gap interface is located at $x_2=30L_{so}$ both with softness parameters $s_1=s_2=0.2L_{so}$. The rest of parameters are $V_0=2E_{so}$, $\Delta_o=0.25E_{so}$,  $\Delta_B=0.4E_{so}$ and $\mu=0$. The function $\mathcal{F}$ is shown in black while the number
of propagating modes in the superconductor side of the junction is shown in blue-gray. Each step corresponds to the activation of a propagating mode. The zeros of $\mathcal{F}$ indicate the existence of a solution with the corresponding energy $E$. b) Same as a) but with a  superconductor gap interface softness $s_2=10L_{so}$ while the potential softness is $s_1=0.2L_{so}$.  c) same as a) and b) but this time with a potential interface softness $s_1=10L_{so}$ and a superconductor gap interface softness $s_2=0.2L_{so}$.}
\label{F5_2}
\end{figure}

A similar robustness against softness
is found in the junction energy spectrum. 
Figure \ref{F5_2}a shows the spectrum of eigenenergies, containing the MZM at zero energy and its closest excited bound and resonant states at finite energies. 
As before, the location of the eigenenergies is signaled by the zeros of the function $\mathcal{F}$ (in black). Note that although the function $\mathcal{F}$ is not symmetrical with respect to $E=0$, the position of the zeros indeed is. 
The particular shape of $\mathcal{F}$ is actually irrelevant and only the position 
of its
zeros bears a physical meaning.
The blue 
staircase curve informs us about the number of propagating modes in the superconductor side of the junction. 
The protection of the MZM, proportional to the energy gap with its
nearby eigenenergies, does not change significantly for moderate values of 
the softness of the superconductor and potential interfaces. 
On the other hand, for high enough values of the softness interesting results arise. 

For high values of the superconductor interface softness, shown in Fig.\ \ref{F5_2}b, the protection of the MZM is increased since its neighboring eigenenergies 
are repelled from zero. In this case, the finite energy modes get closer to the 
activation energy of the propagating modes, i.e., 
to the energy gap  on the superconductor side of the junction.  
On the contrary, the increase of the potential softness introduces more excited states inside the superconductor energy gap, thus getting
closer to the MZM energy (see Fig.\ \ref{F5_2}c). The appearance of 
low energy states in a soft potential interface is in agreement with the results
of Ref.\ \onlinecite{Kells}. The characteristic features of these low energy states
in tunneling conductance experiments were discussed in Ref.\ \onlinecite{StanescuLES}.

When both interfaces are made soft the two effects
on the spectrum we have just discussed compete. That is, the higher softness of the potential introduces
more bound states inside the superconductor energy gap, while the softness of the superconducting interface tries to push them apart from the MZM. The result is 
that many excited states get densely packed near the superconducting gap energy
(see Fig. \ref{F6}b).

\subsection{MZM's in different kinds of junctions}

\begin{figure}[t]
\centering
\includegraphics[width=8.5cm]{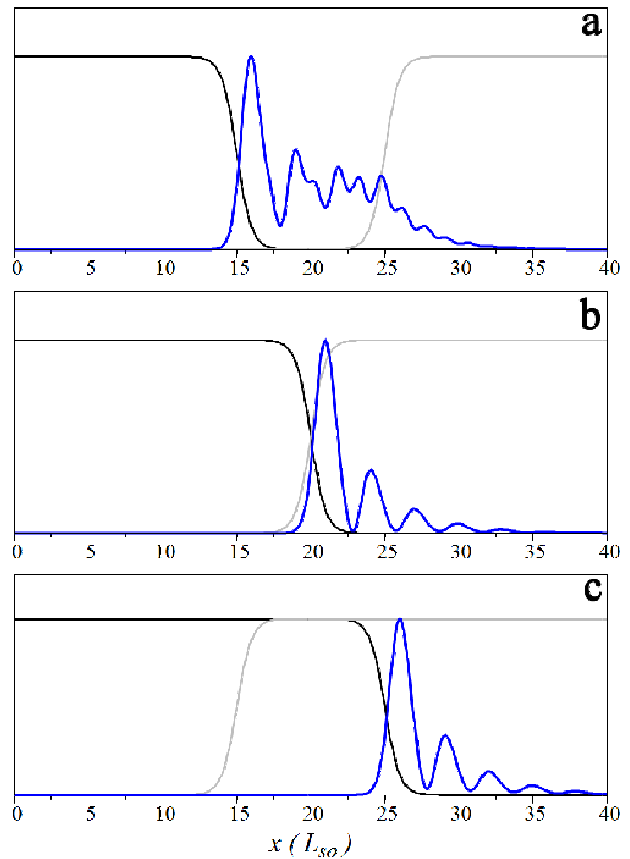}
\caption{ 
(Color online)
Different kinds of soft NS junctions in an infinite nanowire. 
The used parameters are $V_o=2E_{so}$, $\Delta_o=0.25E_{so}$, $\Delta_B=0.4E_{so}$ and $\mu=0.1E_{so}$. Panel a) is for a 
type I junction, with non overlapping high potential and superconductivity regions. Panel  c), for a type II junction, is just the opposite. 
Panel b) is the limiting case
between the two when the potential and superconductor interfaces are located at the same point.}
\label{F7_1}
\end{figure}

Up to this point it has been assumed that the position of the 
potential interface $x_1$ and the superconduction interface $x_2$ 
are such that $x_1<x_2$, i.e., they
do not overlap. In this subsection we consider a more general 
situation, defining two kind of junctions: type I junctions without overlapping 
region ($x_1<x_2$) and type II junctions in the opposite case ($x_1>x_2$). 
Figure \ref{F7_1} shows a comparison between both types, as well as
the limiting intermediate situation. In type I junctions the MZM density behaves as in previous sections, with a density peak localized on the potential edge followed by regular oscillations and a decaying behavior inside the superconductor region. On the other hand, type II junctions just show an oscillatory density whose amplitude decays as the function penetrates the superconductor region. The limiting case $x_1=x_2$ behaves similarly to the type II junction.

We also 
notice from Fig.\ \ref{F7_1} that the density peak is always found 
close to the potential interface. That is, the MZM is located on 
the potential step and not on the superconductivity interface. 
Superconductivity is a
necessary ingredient for the formation of the MZM but, in practice,
its maximum probability can be located quite far from
the superconductor interface.

\begin{figure}[t]
\centering
\includegraphics[width=8.5cm]{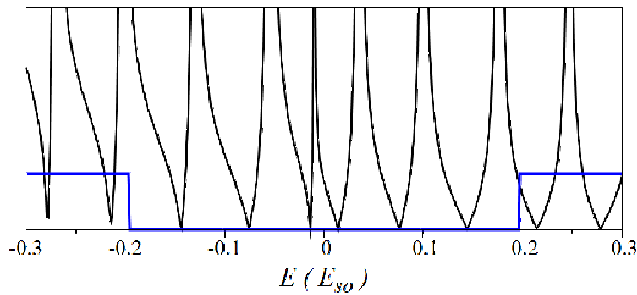}
\caption{(Color  online) Junction spectrum out of the topological phase. In this regime the MZM is split into two subgap fermions. The used parameters are $V_o=2E_{so}$, $\Delta_o=0.6E_{so}$, $\Delta_B=0.4E_{so}$ and $\mu=0$.}
\label{F7_3}
\end{figure}

\begin{figure}[t]
\centering
\includegraphics[width=8.5cm]{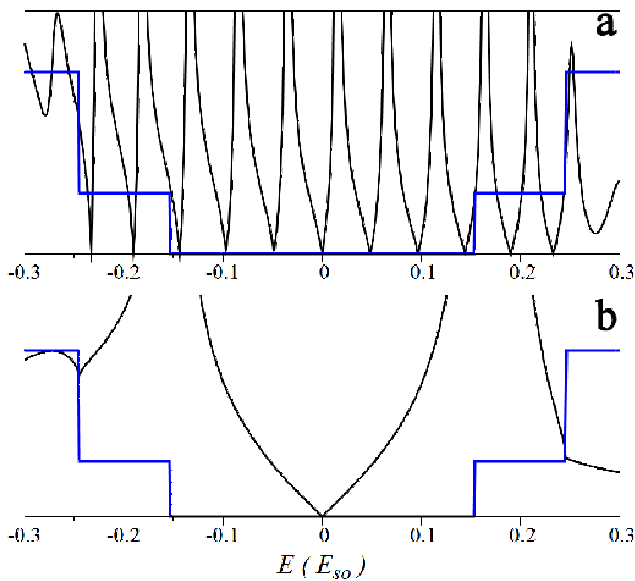}
\caption{
(Color online)
Spectra for different kinds of junctions. 
The used parameters are $V_o=2E_{so}$, $\Delta_o=0.25E_{so}$, $\Delta_B=0.4E_{so}$ and $\mu=0$. Panel a) is for a  type I junction with a long separation between potential and superconductor interfaces ($d=30L_{so}$)
while panel b) corresponds to the limiting case between type I and type II junctions when both interfaces are located on the same position ($d=0$). 
Like in preceding figures the zeros of $\mathcal{F}$ indicate the existence of a solution with the corresponding energy $E$.}
\label{F7_2}
\end{figure}

As shown in Fig. \ref{F5_2}a, bounded states are found in type I junctions at energies different from zero. We believe
these states are Andreev resonant states formed in the region between the 
two interfaces. 
This statement
is confirmed by means of a change in the superconductor bulk value $\Delta_0$. 
As shown in Fig.\ \ref{F7_3}, out of the topological regime the MZM splits into two subgap Fermionic states but the Andreev resonant states remain almost with the same eigenenergies. Notice also that the number of Andreev states is larger and their energies are closer to zero in type I junctions with a large non overlapping region,
i.e., large $d=x_2-x_1$ (see Fig.\ \ref{F7_2}a). On the contrary, if $d$ is diminished the number of Andreev resonant states diminishes an their energies fall apart from zero. In the limiting case when $d$ is zero the Andreev resonant states disappear and the protection of the MZM is determined by the amplitude of the gap on the superconductor side of the junction as shown in Fig \ref{F7_2}b. The same happens for type II junctions with $d<0$. Furthermore, in this case ($d\leq0$) the junction spectrum is even more resilient to changes in the softness of the interfaces, being almost insensitive to them.

\section{Results with input flux}

Decreasing the potential $V_0$, for a fixed $\Delta_B$, $\Delta_0$, and 
zero energy, wave functions characterized by real wave numbers arise in the bulk normal side of the junction. When this occurs, bounded MZM's 
no longer exist due to their coupling with propagating modes.
In the preceding section we assumed that if propagating modes
were present they only carried outgoing flux.
In this section we explore the influence of incident fluxes on the junction.
The same numerical method explained above can be used here,
disregarding the use of the matching point and just fixing the coefficients $C_k$ of the input modes as this already yields a non homogenous linear 
system. 
We only consider input modes from the normal side of the junction, given by 
electron states of positive  $k$ and hole states of  negative $k$. 
Furthermore, it is also assumed that all propagating input modes 
impinge on the junction with exactly the same flux.

\begin{figure}[t]
\centering
\includegraphics[width=8.5cm]{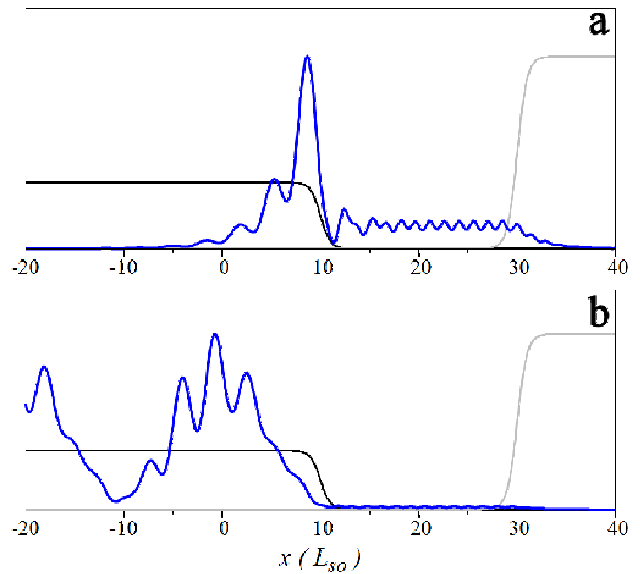}
\caption{
(Color online)
a) Majorana bounded state found for $V_{o}=0.69E_{so}$, $\Delta_{o}=0.25E_{so}$, $\Delta_{B}=0.4E_{so}$ and $\mu=0.1E_{so}$. Potential and superconductor interfaces are located at $x_1=10L_{so}$ and $x_2=30L_{so}$ respectively and their softness parameters are $s_1=s_2=0.5L_{so}$. b) Zero energy non Majorana extended state. The figure is shown for the same parameters as in a) but with $V_{0}=0.67E_{so}$.}
\label{F8}
\end{figure}

Following the sequence from high to low values of $V_0$,
the system evolves from no propagating modes 
at high $V_0$ to four input modes (with two different $k$'s) 
for moderately low values of the potential $V_0$. 
In this case, the resulting zero mode density is characterized by 
a beating pattern of a large wave length modulated by a smaller one 
(see Fig.\ \ref{F8}b). 
For $\Delta_0=0.25E_{so}$ and $\Delta_B=0.4E_{so}$ this regime ranges from $V_0=0.68E_{so}$, where the propagating modes arise, down to $V_0=0.50E_{so}$. Above $V_0=0.68E_{so}$ only evanescent modes are possible (see Fig.\ \ref{F8}a). 
The zero mode solution obtained in this range does not represent a MZM 
since its wave function components do not fulfill the requirement
\begin{equation}
\label{cond}
\Psi_{s_\sigma s_\tau}(x)= (-1)^{\frac{s_\sigma-s_\tau}{2}}\,
\Psi_{-s_\sigma -s_\tau}^*(x)\; .
\end{equation}

\begin{figure}[t]
\centering
\includegraphics[width=8.5cm]{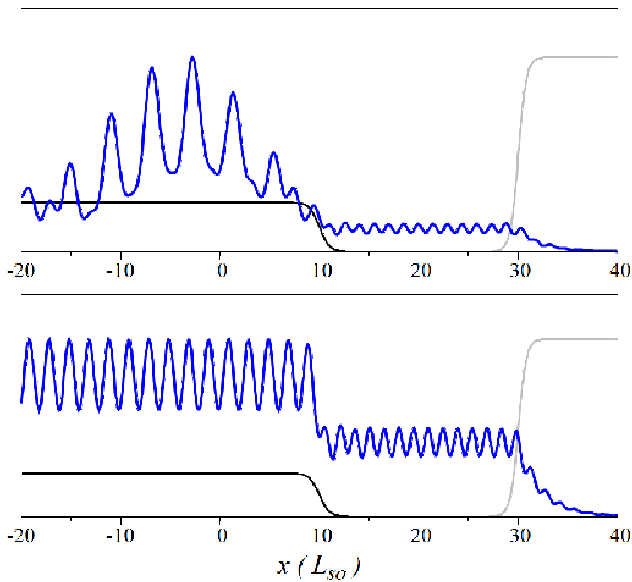}
\caption{
(Color online)
a) Non Majorana bounded state found for $V_0=0.51E_{so}$, $\Delta_{o}=0.25E_{so}$, $\Delta_{B}=0.4E_{so}$ and $\mu=0.1E_{so}$.
Excluding $V_0$ these are the same parameters as in Figs.\ \ref{F8}a and \ref{F8}b. b) 
Density for a 
MZM extended state. The figure is shown for the same parameters as in a) but with $V_{0}=0.49E_{so}$.}
\label{F9}
\end{figure}

For $V_0<0.50 E_{so}$ half of the normal side allowed wave numbers become purely imaginary, thus leading to an evanescent contribution to the boundary condition. As a consequence, there are only two modes (with the same $k$) 
and
the resulting density has a single period of oscillation (see Fig.\ \ref{F9}b). In this case the wave function 
represents a MZM since it fulfills
Eq.\ (\ref{cond}). This example of extended 
MZM's demonstrates that their existence is not limited to bounded states. 
For these extended states the assumption of equal incident flux in 
electron and
hole channels is crucial. If 
the input is prepared in a specific electron or hole state
of a given spin,
the MZM condition Eq.\ (\ref{cond}) is lost 
for low values of the bulk potential. 
Therefore, propagating MZM's are possible albeit  
for a particular superposition of input states only.

\section{Conclusions}
A numerical method to calculate the wave function of MZM's 
in presence of a soft normal-superconductor junction has been 
developed. This method is able to detect whether a particular 
energy $E$ is an eigenenergy or not of the junction 
and, when it is the case, obtain the corresponding wave function.
The junction is described by smooth functions of position in a
1D nanowire with Rashba spin orbit interaction, a Zeeman magnetic field and superconductivity. It has been applied to a semi-infinite 
abrupt nanowire junction in order to compare its results with those obtained analytically, as well as to infinite soft junctions in order to study the dependence to different parameters of the MZM density and its protection from energetically alike excited states. 

We have proven the resilience of the MZM density to the softness parameters 
and studied the dependence of its localization with the potential interface position. This latter feature hints the possibility of manipulating the position of the Majorana modes in order to perform topological quantum operations. We have found the remarkable result of an increase in the protection of the MZM for high values of the softness of the superconductor gap interface, while high values of the softness of the potential interface have an opposite effect. Finally, we have shown the existence of extended MZM's, albeit limited to feed the junction with a particular set of propagating input states. This result demonstrates that 
MZM's are not always restricted to bounded states.

\begin{acknowledgments}
This work was funded by MINECO-Spain (grant FIS2011-23526),
CAIB-Spain (Conselleria d'Educaci\'o, Cultura i Universitats) and 
FEDER. Discussions with R. L\'opez are 
gratefully acknowledged.
 \end{acknowledgments}


\begin{thebibliography}{37}%
\makeatletter
\providecommand \@ifxundefined [1]{%
 \@ifx{#1\undefined}
}%
\providecommand \@ifnum [1]{%
 \ifnum #1\expandafter \@firstoftwo
 \else \expandafter \@secondoftwo
 \fi
}%
\providecommand \@ifx [1]{%
 \ifx #1\expandafter \@firstoftwo
 \else \expandafter \@secondoftwo
 \fi
}%
\providecommand \natexlab [1]{#1}%
\providecommand \enquote  [1]{``#1''}%
\providecommand \bibnamefont  [1]{#1}%
\providecommand \bibfnamefont [1]{#1}%
\providecommand \citenamefont [1]{#1}%
\providecommand \href@noop [0]{\@secondoftwo}%
\providecommand \href [0]{\begingroup \@sanitize@url \@href}%
\providecommand \@href[1]{\@@startlink{#1}\@@href}%
\providecommand \@@href[1]{\endgroup#1\@@endlink}%
\providecommand \@sanitize@url [0]{\catcode `\\12\catcode `\$12\catcode
  `\&12\catcode `\#12\catcode `\^12\catcode `\_12\catcode `\%12\relax}%
\providecommand \@@startlink[1]{}%
\providecommand \@@endlink[0]{}%
\providecommand \url  [0]{\begingroup\@sanitize@url \@url }%
\providecommand \@url [1]{\endgroup\@href {#1}{\urlprefix }}%
\providecommand \urlprefix  [0]{URL }%
\providecommand \Eprint [0]{\href }%
\providecommand \doibase [0]{http://dx.doi.org/}%
\providecommand \selectlanguage [0]{\@gobble}%
\providecommand \bibinfo  [0]{\@secondoftwo}%
\providecommand \bibfield  [0]{\@secondoftwo}%
\providecommand \translation [1]{[#1]}%
\providecommand \BibitemOpen [0]{}%
\providecommand \bibitemStop [0]{}%
\providecommand \bibitemNoStop [0]{.\EOS\space}%
\providecommand \EOS [0]{\spacefactor3000\relax}%
\providecommand \BibitemShut  [1]{\csname bibitem#1\endcsname}%
\let\auto@bib@innerbib\@empty
\bibitem [{\citenamefont {Majorana}(1937)}]{Majorana}%
  \BibitemOpen
  \bibfield  {author} {\bibinfo {author} {\bibfnamefont {E.}~\bibnamefont
  {Majorana}},\ }\href@noop {} {\bibfield  {journal} {\bibinfo  {journal}
  {Nuovo Cimento}\ }\textbf {\bibinfo {volume} {14}},\ \bibinfo {pages} {171}
  (\bibinfo {year} {1937})}\BibitemShut {NoStop}%
\bibitem [{\citenamefont {Kitaev}(2001)}]{Kitaev}%
  \BibitemOpen
  \bibfield  {author} {\bibinfo {author} {\bibfnamefont {A.~Y.}\ \bibnamefont
  {Kitaev}},\ }\href@noop {} {\bibfield  {journal} {\bibinfo  {journal}
  {Physics Uspekhi}\ }\textbf {\bibinfo {volume} {44}},\ \bibinfo {pages} {131}
  (\bibinfo {year} {2001})}\BibitemShut {NoStop}%
\bibitem [{\citenamefont {Wilceck}(2009)}]{Wilczeck}%
  \BibitemOpen
  \bibfield  {author} {\bibinfo {author} {\bibfnamefont {F.}~\bibnamefont
  {Wilceck}},\ }\href@noop {} {\bibfield  {journal} {\bibinfo  {journal}
  {Nature Phys.}\ }\textbf {\bibinfo {volume} {5}},\ \bibinfo {pages} {614}
  (\bibinfo {year} {2009})}\BibitemShut {NoStop}%
\bibitem [{\citenamefont {Alicea}(2012)}]{Alicea}%
  \BibitemOpen
  \bibfield  {author} {\bibinfo {author} {\bibfnamefont {J.}~\bibnamefont
  {Alicea}},\ }\href@noop {} {\bibfield  {journal} {\bibinfo  {journal} {Rep.\
  Prog.\ Phys.}\ }\textbf {\bibinfo {volume} {75}},\ \bibinfo {pages} {076501}
  (\bibinfo {year} {2012})}\BibitemShut {NoStop}%
\bibitem [{\citenamefont {Qi}\ and\ \citenamefont {Zhang}(2011)}]{Qi}%
  \BibitemOpen
  \bibfield  {author} {\bibinfo {author} {\bibfnamefont {X.~L.}\ \bibnamefont
  {Qi}}\ and\ \bibinfo {author} {\bibfnamefont {S.~C.}\ \bibnamefont {Zhang}},\
  }\href@noop {} {\bibfield  {journal} {\bibinfo  {journal} {Rev.\ Mod.\
  Phys.}\ }\textbf {\bibinfo {volume} {83}},\ \bibinfo {pages} {1057} (\bibinfo
  {year} {2011})}\BibitemShut {NoStop}%
\bibitem [{\citenamefont {Leijnse}\ and\ \citenamefont
  {Flensberg}(2012)}]{Leijnse}%
  \BibitemOpen
  \bibfield  {author} {\bibinfo {author} {\bibfnamefont {M.}~\bibnamefont
  {Leijnse}}\ and\ \bibinfo {author} {\bibfnamefont {K.}~\bibnamefont
  {Flensberg}},\ }\href@noop {} {\bibfield  {journal} {\bibinfo  {journal}
  {Semicond.\ Sci.\ Technol.}\ }\textbf {\bibinfo {volume} {27}},\ \bibinfo
  {pages} {124003} (\bibinfo {year} {2012})}\BibitemShut {NoStop}%
\bibitem [{\citenamefont {Beenakker}(2013)}]{Beenakker}%
  \BibitemOpen
  \bibfield  {author} {\bibinfo {author} {\bibfnamefont {C.~W.~J.}\
  \bibnamefont {Beenakker}},\ }\href@noop {} {\bibfield  {journal} {\bibinfo
  {journal} {Annu.\ Rev.\ Condens.\ Matter Phys.}\ }\textbf {\bibinfo {volume}
  {4}},\ \bibinfo {pages} {113} (\bibinfo {year} {2013})}\BibitemShut {NoStop}%
\bibitem [{\citenamefont {Stanescu}\ and\ \citenamefont
  {Tewari}(2013{\natexlab{a}})}]{StanescuREV}%
  \BibitemOpen
  \bibfield  {author} {\bibinfo {author} {\bibfnamefont {T.~D.}\ \bibnamefont
  {Stanescu}}\ and\ \bibinfo {author} {\bibfnamefont {S.}~\bibnamefont
  {Tewari}},\ }\href@noop {} {\bibfield  {journal} {\bibinfo  {journal} {J.
  Phys.\ Condens.\ Matter}\ }\textbf {\bibinfo {volume} {25}},\ \bibinfo
  {pages} {233201} (\bibinfo {year} {2013}{\natexlab{a}})}\BibitemShut
  {NoStop}%
\bibitem [{\citenamefont {Fu}\ and\ \citenamefont {Kane}(2008)}]{Fu}%
  \BibitemOpen
  \bibfield  {author} {\bibinfo {author} {\bibfnamefont {L.}~\bibnamefont
  {Fu}}\ and\ \bibinfo {author} {\bibfnamefont {C.~L.}\ \bibnamefont {Kane}},\
  }\href@noop {} {\bibfield  {journal} {\bibinfo  {journal} {Phys.\ Rev.\
  Lett.}\ }\textbf {\bibinfo {volume} {100}},\ \bibinfo {pages} {096407}
  (\bibinfo {year} {2008})}\BibitemShut {NoStop}%
\bibitem [{\citenamefont {Akhmerov}\ \emph {et~al.}(2009)\citenamefont
  {Akhmerov}, \citenamefont {Nilsson},\ and\ \citenamefont
  {Beenakker}}]{Akhmerov}%
  \BibitemOpen
  \bibfield  {author} {\bibinfo {author} {\bibfnamefont {A.~R.}\ \bibnamefont
  {Akhmerov}}, \bibinfo {author} {\bibfnamefont {J.}~\bibnamefont {Nilsson}}, \
  and\ \bibinfo {author} {\bibfnamefont {C.~W.~J.}\ \bibnamefont {Beenakker}},\
  }\href@noop {} {\bibfield  {journal} {\bibinfo  {journal} {Phys.\ Rev.\
  Lett.}\ }\textbf {\bibinfo {volume} {102}},\ \bibinfo {pages} {216404}
  (\bibinfo {year} {2009})}\BibitemShut {NoStop}%
\bibitem [{\citenamefont {Tanaka}\ \emph {et~al.}(2009)\citenamefont {Tanaka},
  \citenamefont {Yokoyama},\ and\ \citenamefont {Nagaosa}}]{Tanaka}%
  \BibitemOpen
  \bibfield  {author} {\bibinfo {author} {\bibfnamefont {Y.}~\bibnamefont
  {Tanaka}}, \bibinfo {author} {\bibfnamefont {T.}~\bibnamefont {Yokoyama}}, \
  and\ \bibinfo {author} {\bibfnamefont {N.}~\bibnamefont {Nagaosa}},\
  }\href@noop {} {\bibfield  {journal} {\bibinfo  {journal} {Phys.\ Rev.\
  Lett.}\ }\textbf {\bibinfo {volume} {103}},\ \bibinfo {pages} {107002}
  (\bibinfo {year} {2009})}\BibitemShut {NoStop}%
\bibitem [{\citenamefont {Law}\ \emph {et~al.}(2009)\citenamefont {Law},
  \citenamefont {Lee},\ and\ \citenamefont {Ng}}]{Law}%
  \BibitemOpen
  \bibfield  {author} {\bibinfo {author} {\bibfnamefont {K.~T.}\ \bibnamefont
  {Law}}, \bibinfo {author} {\bibfnamefont {P.~A.}\ \bibnamefont {Lee}}, \ and\
  \bibinfo {author} {\bibfnamefont {T.~K.}\ \bibnamefont {Ng}},\ }\href@noop {}
  {\bibfield  {journal} {\bibinfo  {journal} {Phys.\ Rev.\ Lett.}\ }\textbf
  {\bibinfo {volume} {103}},\ \bibinfo {pages} {237001} (\bibinfo {year}
  {2009})}\BibitemShut {NoStop}%
\bibitem [{\citenamefont {Lutchyn}\ \emph {et~al.}(2010)\citenamefont
  {Lutchyn}, \citenamefont {Sau},\ and\ \citenamefont {{Das Sarma}}}]{Lutchyn}%
  \BibitemOpen
  \bibfield  {author} {\bibinfo {author} {\bibfnamefont {R.~M.}\ \bibnamefont
  {Lutchyn}}, \bibinfo {author} {\bibfnamefont {J.~D.}\ \bibnamefont {Sau}}, \
  and\ \bibinfo {author} {\bibfnamefont {S.}~\bibnamefont {{Das Sarma}}},\
  }\href@noop {} {\bibfield  {journal} {\bibinfo  {journal} {Phys.\ Rev.\
  Lett.}\ }\textbf {\bibinfo {volume} {105}},\ \bibinfo {pages} {077001}
  (\bibinfo {year} {2010})}\BibitemShut {NoStop}%
\bibitem [{\citenamefont {Oreg}\ \emph {et~al.}(2010)\citenamefont {Oreg},
  \citenamefont {Refael},\ and\ \citenamefont {von Oppen}}]{Oreg}%
  \BibitemOpen
  \bibfield  {author} {\bibinfo {author} {\bibfnamefont {Y.}~\bibnamefont
  {Oreg}}, \bibinfo {author} {\bibfnamefont {G.}~\bibnamefont {Refael}}, \ and\
  \bibinfo {author} {\bibfnamefont {F.}~\bibnamefont {von Oppen}},\ }\href@noop
  {} {\bibfield  {journal} {\bibinfo  {journal} {Phys.\ Rev.\ Lett.}\ }\textbf
  {\bibinfo {volume} {105}},\ \bibinfo {pages} {177002} (\bibinfo {year}
  {2010})}\BibitemShut {NoStop}%
\bibitem [{\citenamefont {Stanescu}\ \emph {et~al.}(2011)\citenamefont
  {Stanescu}, \citenamefont {Lutchyn},\ and\ \citenamefont {{Das
  Sarma}}}]{Stanescu}%
  \BibitemOpen
  \bibfield  {author} {\bibinfo {author} {\bibfnamefont {T.~D.}\ \bibnamefont
  {Stanescu}}, \bibinfo {author} {\bibfnamefont {R.~M.}\ \bibnamefont
  {Lutchyn}}, \ and\ \bibinfo {author} {\bibfnamefont {S.}~\bibnamefont {{Das
  Sarma}}},\ }\href@noop {} {\bibfield  {journal} {\bibinfo  {journal} {Phys.\
  Rev.\ B}\ }\textbf {\bibinfo {volume} {84}},\ \bibinfo {pages} {144522}
  (\bibinfo {year} {2011})}\BibitemShut {NoStop}%
\bibitem [{\citenamefont {Flensberg}(2010)}]{Flensberg}%
  \BibitemOpen
  \bibfield  {author} {\bibinfo {author} {\bibfnamefont {K.}~\bibnamefont
  {Flensberg}},\ }\href@noop {} {\bibfield  {journal} {\bibinfo  {journal}
  {Phys.\ Rev.\ B}\ }\textbf {\bibinfo {volume} {82}},\ \bibinfo {pages}
  {180516} (\bibinfo {year} {2010})}\BibitemShut {NoStop}%
\bibitem [{\citenamefont {Potter}\ and\ \citenamefont {Lee}(2010)}]{Potter}%
  \BibitemOpen
  \bibfield  {author} {\bibinfo {author} {\bibfnamefont {A.~C.}\ \bibnamefont
  {Potter}}\ and\ \bibinfo {author} {\bibfnamefont {P.~A.}\ \bibnamefont
  {Lee}},\ }\href@noop {} {\bibfield  {journal} {\bibinfo  {journal} {Phys.\
  Rev.\ Lett.}\ }\textbf {\bibinfo {volume} {105}},\ \bibinfo {pages} {227003}
  (\bibinfo {year} {2010})}\BibitemShut {NoStop}%
\bibitem [{\citenamefont {Potter}\ and\ \citenamefont {Lee}(2011)}]{Potter2}%
  \BibitemOpen
  \bibfield  {author} {\bibinfo {author} {\bibfnamefont {A.~C.}\ \bibnamefont
  {Potter}}\ and\ \bibinfo {author} {\bibfnamefont {P.~A.}\ \bibnamefont
  {Lee}},\ }\href@noop {} {\bibfield  {journal} {\bibinfo  {journal} {Phys.\
  Rev.\ B}\ }\textbf {\bibinfo {volume} {83}},\ \bibinfo {pages} {094525}
  (\bibinfo {year} {2011})}\BibitemShut {NoStop}%
\bibitem [{\citenamefont {Gangadharaiah}\ \emph {et~al.}(2011)\citenamefont
  {Gangadharaiah}, \citenamefont {Braunecker}, \citenamefont {Simon},\ and\
  \citenamefont {Loss}}]{Ganga}%
  \BibitemOpen
  \bibfield  {author} {\bibinfo {author} {\bibfnamefont {S.}~\bibnamefont
  {Gangadharaiah}}, \bibinfo {author} {\bibfnamefont {B.}~\bibnamefont
  {Braunecker}}, \bibinfo {author} {\bibfnamefont {P.}~\bibnamefont {Simon}}, \
  and\ \bibinfo {author} {\bibfnamefont {D.}~\bibnamefont {Loss}},\ }\href@noop
  {} {\bibfield  {journal} {\bibinfo  {journal} {Phys.\ Rev.\ Lett.}\ }\textbf
  {\bibinfo {volume} {107}},\ \bibinfo {pages} {036801} (\bibinfo {year}
  {2011})}\BibitemShut {NoStop}%
\bibitem [{\citenamefont {Egger}\ \emph {et~al.}(2010)\citenamefont {Egger},
  \citenamefont {Zazunov},\ and\ \citenamefont {Yeyati}}]{Egger}%
  \BibitemOpen
  \bibfield  {author} {\bibinfo {author} {\bibfnamefont {R.}~\bibnamefont
  {Egger}}, \bibinfo {author} {\bibfnamefont {A.}~\bibnamefont {Zazunov}}, \
  and\ \bibinfo {author} {\bibfnamefont {A.~L.}\ \bibnamefont {Yeyati}},\
  }\href@noop {} {\bibfield  {journal} {\bibinfo  {journal} {Phys.\ Rev.\
  Lett.}\ }\textbf {\bibinfo {volume} {105}},\ \bibinfo {pages} {136403}
  (\bibinfo {year} {2010})}\BibitemShut {NoStop}%
\bibitem [{\citenamefont {Zazunov}\ \emph {et~al.}(2011)\citenamefont
  {Zazunov}, \citenamefont {Yeyati},\ and\ \citenamefont {Egger}}]{Zazunov}%
  \BibitemOpen
  \bibfield  {author} {\bibinfo {author} {\bibfnamefont {A.}~\bibnamefont
  {Zazunov}}, \bibinfo {author} {\bibfnamefont {A.~L.}\ \bibnamefont {Yeyati}},
  \ and\ \bibinfo {author} {\bibfnamefont {R.}~\bibnamefont {Egger}},\
  }\href@noop {} {\bibfield  {journal} {\bibinfo  {journal} {Phys.\ Rev.\ B}\
  }\textbf {\bibinfo {volume} {84}},\ \bibinfo {pages} {165440} (\bibinfo
  {year} {2011})}\BibitemShut {NoStop}%
\bibitem [{\citenamefont {Klinovaja}\ \emph {et~al.}(2012)\citenamefont
  {Klinovaja}, \citenamefont {Gangadharaiah},\ and\ \citenamefont
  {Loss}}]{Klino1}%
  \BibitemOpen
  \bibfield  {author} {\bibinfo {author} {\bibfnamefont {J.}~\bibnamefont
  {Klinovaja}}, \bibinfo {author} {\bibfnamefont {S.}~\bibnamefont
  {Gangadharaiah}}, \ and\ \bibinfo {author} {\bibfnamefont {D.}~\bibnamefont
  {Loss}},\ }\href@noop {} {\bibfield  {journal} {\bibinfo  {journal} {Phys.\
  Rev.\ Lett.}\ }\textbf {\bibinfo {volume} {108}},\ \bibinfo {pages} {196804}
  (\bibinfo {year} {2012})}\BibitemShut {NoStop}%
\bibitem [{\citenamefont {Klinovaja}\ and\ \citenamefont
  {Loss}(2012)}]{Klino2}%
  \BibitemOpen
  \bibfield  {author} {\bibinfo {author} {\bibfnamefont {J.}~\bibnamefont
  {Klinovaja}}\ and\ \bibinfo {author} {\bibfnamefont {D.}~\bibnamefont
  {Loss}},\ }\href@noop {} {\bibfield  {journal} {\bibinfo  {journal} {Phys.\
  Rev.\ B}\ }\textbf {\bibinfo {volume} {86}},\ \bibinfo {pages} {085408}
  (\bibinfo {year} {2012})}\BibitemShut {NoStop}%
\bibitem [{\citenamefont {Lim}\ \emph {et~al.}(2012{\natexlab{a}})\citenamefont
  {Lim}, \citenamefont {Serra}, \citenamefont {Lopez},\ and\ \citenamefont
  {Aguado}}]{Lim}%
  \BibitemOpen
  \bibfield  {author} {\bibinfo {author} {\bibfnamefont {J.~S.}\ \bibnamefont
  {Lim}}, \bibinfo {author} {\bibfnamefont {L.}~\bibnamefont {Serra}}, \bibinfo
  {author} {\bibfnamefont {R.}~\bibnamefont {Lopez}}, \ and\ \bibinfo {author}
  {\bibfnamefont {R.}~\bibnamefont {Aguado}},\ }\href@noop {} {\bibfield
  {journal} {\bibinfo  {journal} {Phys.\ Rev.\ B}\ }\textbf {\bibinfo {volume}
  {86}},\ \bibinfo {pages} {121103} (\bibinfo {year}
  {2012}{\natexlab{a}})}\BibitemShut {NoStop}%
\bibitem [{\citenamefont {Lim}\ \emph {et~al.}(2012{\natexlab{b}})\citenamefont
  {Lim}, \citenamefont {Lopez},\ and\ \citenamefont {Serra}}]{Lim2}%
  \BibitemOpen
  \bibfield  {author} {\bibinfo {author} {\bibfnamefont {J.~S.}\ \bibnamefont
  {Lim}}, \bibinfo {author} {\bibfnamefont {R.}~\bibnamefont {Lopez}}, \ and\
  \bibinfo {author} {\bibfnamefont {L.}~\bibnamefont {Serra}},\ }\href@noop {}
  {\bibfield  {journal} {\bibinfo  {journal} {New J. Phys.}\ }\textbf {\bibinfo
  {volume} {14}},\ \bibinfo {pages} {083020} (\bibinfo {year}
  {2012}{\natexlab{b}})}\BibitemShut {NoStop}%
\bibitem [{\citenamefont {Lim}\ \emph {et~al.}(2013)\citenamefont {Lim},
  \citenamefont {Lopez},\ and\ \citenamefont {Serra}}]{Lim3}%
  \BibitemOpen
  \bibfield  {author} {\bibinfo {author} {\bibfnamefont {J.~S.}\ \bibnamefont
  {Lim}}, \bibinfo {author} {\bibfnamefont {R.}~\bibnamefont {Lopez}}, \ and\
  \bibinfo {author} {\bibfnamefont {L.}~\bibnamefont {Serra}},\ }\href@noop {}
  {\bibfield  {journal} {\bibinfo  {journal} {Europhys.\ Lett.\ (in press,
  arXiv:1303.7447)}\ } (\bibinfo {year} {2013})}\BibitemShut {NoStop}%
\bibitem [{\citenamefont {Mourik}\ \emph {et~al.}(2012)\citenamefont {Mourik},
  \citenamefont {Zuo}, \citenamefont {Frolov}, \citenamefont {Plissard},
  \citenamefont {Bakkers},\ and\ \citenamefont {Kouwenhoven}}]{Mourik}%
  \BibitemOpen
  \bibfield  {author} {\bibinfo {author} {\bibfnamefont {V.}~\bibnamefont
  {Mourik}}, \bibinfo {author} {\bibfnamefont {K.}~\bibnamefont {Zuo}},
  \bibinfo {author} {\bibfnamefont {S.}~\bibnamefont {Frolov}}, \bibinfo
  {author} {\bibfnamefont {S.}~\bibnamefont {Plissard}}, \bibinfo {author}
  {\bibfnamefont {E.}~\bibnamefont {Bakkers}}, \ and\ \bibinfo {author}
  {\bibfnamefont {L.}~\bibnamefont {Kouwenhoven}},\ }\href@noop {} {\bibfield
  {journal} {\bibinfo  {journal} {Science}\ }\textbf {\bibinfo {volume}
  {336}},\ \bibinfo {pages} {1003} (\bibinfo {year} {2012})}\BibitemShut
  {NoStop}%
\bibitem [{\citenamefont {Deng}\ \emph {et~al.}(2012)\citenamefont {Deng},
  \citenamefont {Yu}, \citenamefont {Huan}, \citenamefont {Larsson},\ and\
  \citenamefont {Caroff}}]{Deng}%
  \BibitemOpen
  \bibfield  {author} {\bibinfo {author} {\bibfnamefont {M.~T.}\ \bibnamefont
  {Deng}}, \bibinfo {author} {\bibfnamefont {C.~L.}\ \bibnamefont {Yu}},
  \bibinfo {author} {\bibfnamefont {G.~Y.}\ \bibnamefont {Huan}}, \bibinfo
  {author} {\bibfnamefont {M.}~\bibnamefont {Larsson}}, \ and\ \bibinfo
  {author} {\bibfnamefont {P.}~\bibnamefont {Caroff}},\ }\href@noop {}
  {\bibfield  {journal} {\bibinfo  {journal} {Nano Lett.}\ }\textbf {\bibinfo
  {volume} {12}},\ \bibinfo {pages} {6414} (\bibinfo {year}
  {2012})}\BibitemShut {NoStop}%
\bibitem [{\citenamefont {Rokhinson}\ \emph {et~al.}(2012)\citenamefont
  {Rokhinson}, \citenamefont {Liu},\ and\ \citenamefont {Furdyna}}]{Rokhinson}%
  \BibitemOpen
  \bibfield  {author} {\bibinfo {author} {\bibfnamefont {L.~P.}\ \bibnamefont
  {Rokhinson}}, \bibinfo {author} {\bibfnamefont {X.}~\bibnamefont {Liu}}, \
  and\ \bibinfo {author} {\bibfnamefont {J.~K.}\ \bibnamefont {Furdyna}},\
  }\href@noop {} {\bibfield  {journal} {\bibinfo  {journal} {Nature Physics}\
  }\textbf {\bibinfo {volume} {8}},\ \bibinfo {pages} {795} (\bibinfo {year}
  {2012})}\BibitemShut {NoStop}%
\bibitem [{\citenamefont {Das}\ \emph {et~al.}(2012)\citenamefont {Das},
  \citenamefont {Ronen}, \citenamefont {Most}, \citenamefont {Oreg},
  \citenamefont {Heiblum},\ and\ \citenamefont {Shtrikman}}]{Das}%
  \BibitemOpen
  \bibfield  {author} {\bibinfo {author} {\bibfnamefont {A.}~\bibnamefont
  {Das}}, \bibinfo {author} {\bibfnamefont {Y.}~\bibnamefont {Ronen}}, \bibinfo
  {author} {\bibfnamefont {Y.}~\bibnamefont {Most}}, \bibinfo {author}
  {\bibfnamefont {Y.}~\bibnamefont {Oreg}}, \bibinfo {author} {\bibfnamefont
  {M.}~\bibnamefont {Heiblum}}, \ and\ \bibinfo {author} {\bibfnamefont
  {H.}~\bibnamefont {Shtrikman}},\ }\href@noop {} {\bibfield  {journal}
  {\bibinfo  {journal} {Nature Physics}\ }\textbf {\bibinfo {volume} {8}},\
  \bibinfo {pages} {887} (\bibinfo {year} {2012})}\BibitemShut {NoStop}%
\bibitem [{\citenamefont {Finck}\ \emph {et~al.}(2013)\citenamefont {Finck},
  \citenamefont {Van~Harlingen}, \citenamefont {Mohseni}, \citenamefont
  {Jung},\ and\ \citenamefont {Li}}]{Finck}%
  \BibitemOpen
  \bibfield  {author} {\bibinfo {author} {\bibfnamefont {A.~D.~K.}\
  \bibnamefont {Finck}}, \bibinfo {author} {\bibfnamefont {D.~J.}\ \bibnamefont
  {Van~Harlingen}}, \bibinfo {author} {\bibfnamefont {P.~K.}\ \bibnamefont
  {Mohseni}}, \bibinfo {author} {\bibfnamefont {K.}~\bibnamefont {Jung}}, \
  and\ \bibinfo {author} {\bibfnamefont {X.}~\bibnamefont {Li}},\ }\href
  {\doibase 10.1103/PhysRevLett.110.126406} {\bibfield  {journal} {\bibinfo
  {journal} {Phys. Rev. Lett.}\ }\textbf {\bibinfo {volume} {110}},\ \bibinfo
  {pages} {126406} (\bibinfo {year} {2013})}\BibitemShut {NoStop}%
\bibitem [{\citenamefont {Pachos}(2012)}]{Pachos}%
  \BibitemOpen
  \bibfield  {author} {\bibinfo {author} {\bibfnamefont {J.~K.}\ \bibnamefont
  {Pachos}},\ }\href@noop {} {\emph {\bibinfo {title} {Introduction to
  topological Quantum Computation}}}\ (\bibinfo  {publisher} {Cambridge
  University Press},\ \bibinfo {year} {2012})\BibitemShut {NoStop}%
\bibitem [{\citenamefont {Nayak}\ \emph {et~al.}(2008)\citenamefont {Nayak},
  \citenamefont {Simon}, \citenamefont {Stern}, \citenamefont {Freedman},\ and\
  \citenamefont {{Das Sarma}}}]{Nayak}%
  \BibitemOpen
  \bibfield  {author} {\bibinfo {author} {\bibfnamefont {C.}~\bibnamefont
  {Nayak}}, \bibinfo {author} {\bibfnamefont {S.~H.}\ \bibnamefont {Simon}},
  \bibinfo {author} {\bibfnamefont {A.}~\bibnamefont {Stern}}, \bibinfo
  {author} {\bibfnamefont {M.}~\bibnamefont {Freedman}}, \ and\ \bibinfo
  {author} {\bibfnamefont {S.}~\bibnamefont {{Das Sarma}}},\ }\href@noop {}
  {\bibfield  {journal} {\bibinfo  {journal} {Rev.\ Mod.\ Phys.}\ }\textbf
  {\bibinfo {volume} {80}},\ \bibinfo {pages} {1083} (\bibinfo {year}
  {2008})}\BibitemShut {NoStop}%
\bibitem [{\citenamefont {Kells}\ \emph {et~al.}(2012)\citenamefont {Kells},
  \citenamefont {Meidan},\ and\ \citenamefont {Brouwer}}]{Kells}%
  \BibitemOpen
  \bibfield  {author} {\bibinfo {author} {\bibfnamefont {G.}~\bibnamefont
  {Kells}}, \bibinfo {author} {\bibfnamefont {D.}~\bibnamefont {Meidan}}, \
  and\ \bibinfo {author} {\bibfnamefont {P.~W.}\ \bibnamefont {Brouwer}},\
  }\href@noop {} {\bibfield  {journal} {\bibinfo  {journal} {Phys.\ Rev.\ B}\
  }\textbf {\bibinfo {volume} {86}},\ \bibinfo {pages} {100503} 
  (\bibinfo {year} {2012})}\BibitemShut
  {NoStop}%
\bibitem [{\citenamefont {Serra}(2013)}]{Serra}%
  \BibitemOpen
  \bibfield  {author} {\bibinfo {author} {\bibfnamefont {L.}~\bibnamefont
  {Serra}},\ }\href@noop {} {\bibfield  {journal} {\bibinfo  {journal} {Phys.\
  Rev.\ B}\ }\textbf {\bibinfo {volume} {87}},\ \bibinfo {pages} {075440}
  (\bibinfo {year} {2013})}\BibitemShut {NoStop}%
\bibitem [{Har()}]{Harwell}%
  \BibitemOpen
  \href@noop {} {}\bibinfo {note} {HSL (2013). A collection of Fortran codes
  for large scale scientific computation. http://www.hsl.rl.ac.uk"}\BibitemShut
  {NoStop}%
\bibitem [{\citenamefont {Stanescu}\ and\ \citenamefont
  {Tewari}(2013{\natexlab{b}})}]{StanescuLES}%
  \BibitemOpen
  \bibfield  {author} {\bibinfo {author} {\bibfnamefont {T.~D.}\ \bibnamefont
  {Stanescu}}\ and\ \bibinfo {author} {\bibfnamefont {S.}~\bibnamefont
  {Tewari}},\ }\href@noop {} {\bibfield  {journal} {\bibinfo  {journal} {Phys.\
  Rev.\ B}\ }\textbf {\bibinfo {volume} {87}},\ \bibinfo {pages} {140504}
  (\bibinfo {year} {2013}{\natexlab{b}})}\BibitemShut {NoStop}%
\end{thebibliography}

%

\end{document}